\documentclass[conference]{IEEEtran}
\IEEEoverridecommandlockouts
\usepackage{cite}
\usepackage{amsmath,amssymb,amsfonts}
\usepackage{graphicx}
\usepackage{textcomp}
\usepackage{booktabs}
\usepackage{threeparttable}
\usepackage{makecell}
\usepackage{url}
\usepackage[ruled,vlined,linesnumbered]{algorithm2e}
\usepackage{array}
\usepackage[table]{xcolor}
\usepackage{pifont}
\usepackage{tabularx}
\usepackage{tikz}
\usetikzlibrary{arrows.meta,shapes.multipart,calc}

\definecolor{c1}{HTML}{0c8918}
\definecolor{c2}{HTML}{dc3023}
\definecolor{darkblue}{rgb}{0, 0.0, 0.78}

\definecolor{tabletext}{RGB}{15, 23, 42}
\definecolor{resultcolor}{RGB}{0, 120, 90}
\definecolor{errorcolor}{RGB}{220, 38, 38}
\definecolor{syntaxcolor}{RGB}{30, 41, 59}
\definecolor{keywordcolor}{RGB}{67, 56, 202}
\definecolor{codebg}{RGB}{243, 244, 246}
\definecolor{commentcolor}{RGB}{51, 65, 85}

\definecolor{darkgreen}{RGB}{0, 102, 51}
\definecolor{purple}{RGB}{102, 0, 153}
\definecolor{lightgray}{RGB}{240, 240, 240}

\def\BibTeX{{\rm B\kern-.05em{\sc i\kern-.025em b}\kern-.08em
    T\kern-.1667em\lower.7ex\hbox{E}\kern-.125emX}}

\newcommand\blfootnote[1]{
  \begingroup
  \renewcommand\thefootnote{}\footnote{#1}
  \addtocounter{footnote}{-1}
  \endgroup
}
\newcommand{\cmark}{\textcolor{green!60!black}{\ding{51}}}
\newcommand{\xmark}{\textcolor{red!80!black}{\ding{55}}}
\newcommand{\cmarkbold}{\textcolor{green!60!black}{\pmb{\ding{51}}}}

\newcolumntype{C}{>{\centering\arraybackslash}X}

\begin{document}

\title{Bridging the Gap: Enabling Natural Language Queries for NoSQL Databases through Text-to-NoSQL Translation}

\author{
  \IEEEauthorblockN{
    Jinwei Lu$^1$,
    Jiawei Lu$^1$,
    Chen Zhang$^1$,
    Zhiqian Qin$^1$,\\
    Haodi Zhang$^2$,
    Yuanfeng Song$^3$,
    Raymond Chi-Wing Wong$^4$
  }
  \IEEEauthorblockA{
    $^1$The Hong Kong Polytechnic University, Hong Kong, China\\
    $^2$Shenzhen University, Shenzhen, China
    \quad
    $^3$WeBank Co., Ltd, Shenzhen, China\\
    $^4$The Hong Kong University of Science and Technology, Hong Kong, China
  }
  \vspace{-1.8em}
}

\maketitle

\blfootnote{A demonstration version of this work, Querycraft, has been accepted to the VLDB 2026 Demo Track\cite{lu2026demonstration}.}

\begin{abstract}
NoSQL databases are core data infrastructure, yet natural-language access to them remains underdeveloped: correct query generation must recover how a non-relational data model represents entities, nested paths, arrays, missing fields, and dynamic keys. This paper studies \textbf{Text-to-NoSQL}, translating natural-language requests into executable NoSQL queries, instantiated with MongoDB aggregation pipelines over schema-less document stores. We present TEND (short for \underline{Te}xt-to-\underline{N}oSQL \underline{D}ataset), an execution-verified benchmark with 1,210 MongoDB-native tasks across 11 databases. To our knowledge, TEND is the first Text-to-NoSQL benchmark whose database worlds are MongoDB-native by design: experts manually define collection boundaries, nested arrays, optional and sparse paths, polymorphic shapes, and dynamic-key conventions; these worlds are populated with real data and verified through frozen MongoDB execution, so TEND evaluates schema-less document reasoning rather than SQL-to-MQL transfer. We further introduce SAG, a \underline{S}chema-\underline{a}s-Data \underline{G}rounding solver that induces path and value grounding from stored-document evidence before bounded MQL generation, execution-grounded repair, and result-consistency selection. Evaluation uses bounded column-tolerant execution accuracy (EXC) as the headline metric, complemented by a graded result-set $F_1$ and a mutually exclusive execution-outcome decomposition. Experiments show that LLMs with strong NL2SQL performance degrade substantially on TEND, validating Text-to-NoSQL as a distinct schema-less document reasoning problem.
\end{abstract}


\section{Introduction}
\label{sec:intro}

\begin{figure*}
    \centering
    \includegraphics[width=0.84\textwidth]{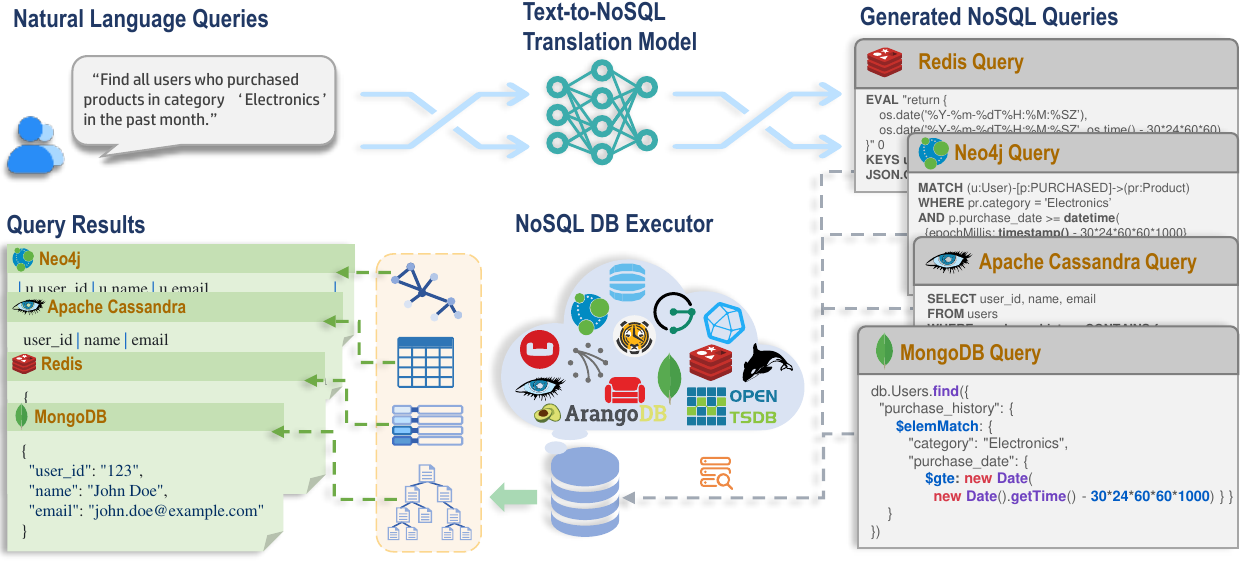}
    \caption{The Text-to-NoSQL problem family: natural-language requests are translated into executable queries across NoSQL data models and query languages (Redis, Neo4j, Cassandra, and MongoDB illustrated); TEND and SAG instantiate the document-store branch with MongoDB aggregation pipelines.}
    \vspace{-10pt}
    \label{fig:text2nosql}
\end{figure*}

NoSQL databases have become a central part of modern data management because they support flexible records, horizontal scaling, and heterogeneous application state without forcing every workload into fixed relational tables, and the database community has studied them from multiple angles, including data modeling, scalability, storage management, and distributed execution \cite{vera2021data,gessert2016scalable,davoudian2018survey,6106531,10.1016/j.future.2015.05.003,6625441,10.1145/1978915.1978919}. NoSQL is an umbrella term: document, graph, key-value, and wide-column systems relax relational assumptions in different ways and expose different query languages. This paper focuses on the document-store branch through MongoDB, a practically consequential representative whose aggregation framework makes schema-less document reasoning central to query generation \cite{dbenginesDocumentStores2026}.

Natural-language database interfaces have made substantial progress for relational databases, where Text-to-SQL can rely on explicit table-column catalogs; the NoSQL setting remains much less systematized, and the gap is deeper than query-language syntax. For MongoDB, correct aggregation pipelines often depend on optional paths, arrays that must be unwound before grouping, keys that are themselves data values, and aliases introduced by earlier pipeline stages. Many of these structures are witnessed by stored documents rather than guaranteed by a fixed catalog, so direct transfer from Text-to-SQL or SQL-to-MQL syntax conversion misses the central database challenge.

We therefore study \textbf{Text-to-NoSQL}: translating a natural-language query (NLQ) into an executable query over a concrete NoSQL database instance (Figure~\ref{fig:text2nosql}). The term names the broader problem family across non-relational data models and target languages; this paper instantiates it with MongoDB aggregation pipelines over schema-less document stores, where the technical problem is executable query synthesis under schema uncertainty, nested structure, dynamic-key evidence, and instance-witnessed semantics rather than MongoDB-like surface syntax.

To support systematic advancement in this field, we release the \textbf{TEND} dataset, which addresses the benchmark gap for MongoDB-native Text-to-NoSQL. Existing resources that touch MongoDB commonly inherit relational regularity by translating schemas or gold SQL into MQL targets; such construction is useful for syntax transfer but does not make document modeling itself the object of evaluation. TEND instead uses BIRD mini-dev \cite{NEURIPS2023_83fc8fab} only as source evidence for realistic domains, values, and workload pressure. Experts then design a new MongoDB DataWorld for each selected domain, deciding collection boundaries, embedded records, array nesting, dynamic-key maps, optional paths, sparse fields, polymorphic shapes, and value witnesses by hand. Each task is paired with an expert-refined MongoDB aggregation pipeline, a canonical English request, and frozen witness data for execution verification; Section~\ref{sec:tend} details the construction protocol and statistics.

The schema-as-data challenge also shapes the solver. We develop \textbf{SAG}, a \underline{S}chema-\underline{a}s-Data \underline{G}rounding solver for Text-to-NoSQL. SAG treats stored documents as the evidence from which the operative schema is induced. Before generation, it performs bounded witness sampling, builds a \texttt{GroundingIndex}, renders a closed path card with dynamic-key abstractions, anchors literals through value witnesses, and then uses bounded decode/repair with execution feedback and result-space consistency to produce a MongoDB aggregation pipeline. The design is intentionally not an open-ended exploratory agent: database grounding is performed deterministically before generation, while the loop is reserved for repairing concrete violations exposed by grounding and execution.

Finally, Text-to-NoSQL needs evaluation that reflects both executable database behavior and diagnosable query-generation failures. We use bounded column-tolerant execution accuracy (\textbf{EXC}) as the headline execution metric because MongoDB aggregation outputs may contain benign helper fields while still needing strict row, value, nested-object, and order semantics. We complement it with a graded result-set $F_1$ (\textbf{EXF\textsubscript{1}}) that measures how close a wrong answer is to the gold result, and with a mutually exclusive outcome decomposition that attributes every failure to a single cause. Extensive experiments show that LLMs that perform strongly on NL2SQL benchmarks degrade sharply on TEND, indicating that TEND is not a surface-language variant of Text-to-SQL but a benchmark for schema-less document reasoning (Section~\ref{sec:exp_analysis}).

The primary contributions of this paper are summarized as follows:
\begin{itemize}
\item We formalize MongoDB-native Text-to-NoSQL as executable aggregation-pipeline synthesis over schema-less document stores, keeping Text-to-NoSQL as the broader problem family.
\item We construct TEND\footnote{The dataset and source code are available at \url{https://github.com/Jinwei-Lu/Text-to-NoSQL}}, to our knowledge the first MongoDB-native, expert-guided, and execution-verified Text-to-NoSQL benchmark whose database worlds are manually designed for schema-less document reasoning. TEND contains 1,210 tasks across 11 databases and makes nesting, dynamic keys, sparse and optional paths, polymorphic shapes, and witness values first-class benchmark variables.
\item We design SAG, a schema-as-data grounding solver that converts stored-document evidence into a path lattice, value witnesses, alignment gates, execution-grounded repair signals, and result-consistency clusters before emitting MQL.
\item We provide an execution-centered evaluation protocol with EXC as the headline metric, a graded result-set $F_1$ (EXF\textsubscript{1}), a mutually exclusive execution-outcome decomposition, and paired significance testing, with every failure retained in the denominator.
\item We set up controlled comparisons against channel-pure direct, sampled-document, relational SQL-pivot, and bounded ReAct baselines, all sharing the same backbone model and evaluator, so differences are attributable to the information channel and mechanism.
\end{itemize}

\section{The TEND Benchmark}
\label{sec:tend}
TEND is a human-annotated, execution-verified benchmark for MongoDB-native Text-to-NoSQL over schema-less document stores. Its construction is design-first: database experts author each MongoDB DataWorld and its tasks from the domain's analytic needs (Section~\ref{sec:dataworld_construction}), and the BIRD mini-dev databases \cite{NEURIPS2023_83fc8fab} serve only as a source of real values, domain semantics, and analytic workload pressure, never as a SQL-to-MQL translation target, schema blueprint, or execution oracle.

\begin{figure*}[htbp]
    \centering
    \includegraphics[width=0.8\textwidth]{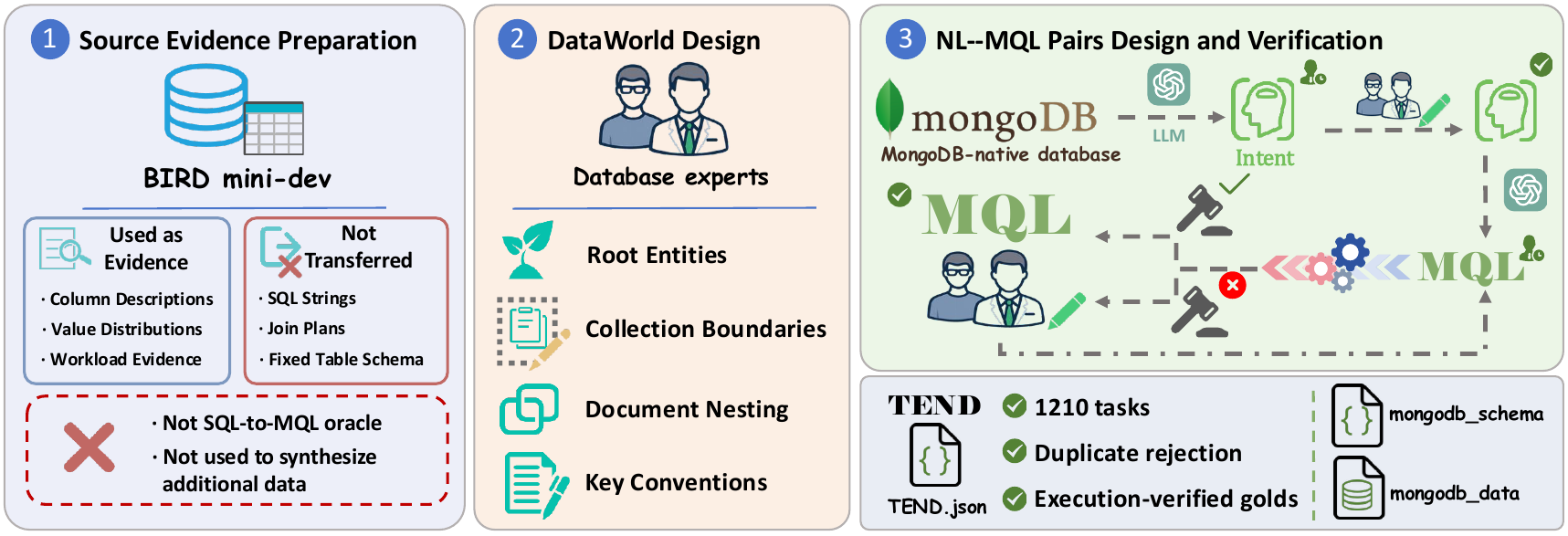}
    \caption{Human annotation pipeline for TEND: seven database experts define one MongoDB-native construction specification per database, LLMs draft MQL programs at scale, and experts refine and execution-verify all 1,210 NL--MQL tasks.}
    \label{fig:tend}
    \vspace{0.5em}
\end{figure*}

\subsection{Core Design of TEND}
\label{sec:tend_scope}

TEND is not a relational benchmark wrapped in MongoDB syntax. Its core objective is to evaluate whether a system can synthesize executable aggregation pipelines over document data whose operative schema is partly encoded in the stored instances themselves. To the best of our knowledge, TEND is the first Text-to-NoSQL benchmark whose databases are MongoDB-native DataWorlds rather than table-to-collection migrations, SQL-equivalent query targets, or generic multi-model translations.

\textbf{Schema-less structure as the target of reasoning.} In TEND, \emph{schema-less} does not mean structure-free: the database provides no global catalog that fully determines every document shape, and TEND turns this flexibility into query semantics. A request may depend on a map whose keys are domain values, an embedded record carrying local evidence, an array that must be unwound before aggregation, or a field whose absence means something different from null, so the operative schema must be inferred from stored-document witnesses.

\textbf{MongoDB-native DataWorlds.} Native means the database design is meaningful as a document store, not merely that data are loaded into MongoDB. Experts decide collection boundaries, embedded entities, nesting depth, dynamic-key conventions, optional paths, sparse fields, polymorphic shapes, and witness structures according to domain analytic needs; BIRD rows supply realistic values, but all structural decisions are manual, making these structures benchmark variables rather than conversion artifacts.

\textbf{MQL-centered difficulty.} Every gold program is authored and verified as a MongoDB aggregation pipeline, not as a syntactic rewrite of a SQL query. Many tasks first build an analytic view of the document world and only then compute the requested answer, with pipelines reaching 13 stages (Section~\ref{sec:tend_statistics}). Intermediate fields, reshaped arrays, dynamic-key expansions, and derived group keys are not incidental implementation details; they are the semantic path from the natural-language request to the final result. TEND therefore tests document-level reasoning and aggregation planning, not field selection over a known schema.

\textbf{Native stressors and exposed failures.} TEND concentrates seven query-bearing document structures, each paired with the failure pattern it exposes in generated MQL: \emph{dynamic-key maps} (domain values stored as object keys) expose solvers that treat keys as fixed columns and lose key--value identity; \emph{nested arrays} expose unwinding at the wrong grain, which duplicates or drops evidence; \emph{optional and sparse paths} expose solvers that assume uniform fields and filter out meaningful absence; \emph{polymorphic shapes} expose commitment to a single document variant while valid alternates are ignored; \emph{embedded records} expose invented joins that flatten away co-located context; \emph{value witnesses} expose literals misbound to the wrong path or stored casing; and \emph{result contracts} expose leaked helper fields, group keys, and sort artifacts. Figure~\ref{fig:tend-example} works one such stressor end to end.

\begin{figure}[t]
\centering
\begin{tikzpicture}[
  x=1pt,
  y=1pt,
  font=\sffamily\footnotesize,
  panel/.style 2 args={
    rectangle split, rectangle split parts=2,
    rectangle split part align={left,left},
    rectangle split part fill={#1,white},
    rectangle split draw splits=true,
    draw=#2, line width=0.5pt,
    inner xsep=5pt, inner ysep=3.5pt,
    text width=228pt, align=left, anchor=north
  },
  flow/.style={-{Latex[length=5.5pt,width=6pt]}, draw=black!70, line width=0.8pt},
  ops/.style={anchor=west, inner xsep=4pt, inner ysep=1pt, align=left,
              font=\fontsize{6.5}{8}\selectfont\ttfamily, text=black!75},
  strip/.style={draw=black!30, line width=0.4pt, rounded corners=1.5pt,
                text width=228pt, inner xsep=5pt, inner ysep=3.2pt,
                align=left, font=\sffamily\scriptsize}
]
\newcommand{\panelcode}[1]{\begin{minipage}[t]{124pt}\fontsize{6.5}{8}\selectfont\ttfamily #1\end{minipage}}
\newcommand{\panelnote}[1]{\begin{minipage}[t]{92pt}\fontsize{6.5}{8.5}\selectfont\sffamily\color{black!58}#1\end{minipage}}
\node[strip, fill=black!4] (nlq) at (0,0) {%
  \textbf{NLQ}\hspace{6pt}For each team and season, compute matches
  played, home wins, away losses, and goal difference; keep team-seasons
  with at least 30 matches, 8 home wins, and 5 away losses; sort by home
  wins descending with seven tie-breakers and return the top 10.%
};
\node[panel={blue!11}{black!45}] (stored) at ($(nlq.south)+(0,-6)$) {%
  \textbf{\strut Stored document}
  \nodepart{two}
  \panelcode{team: \{ long\_name: "Aberdeen" \}\\
  matches\_by\_season: \{\\
  \ "2008/2009": \{ fixtures: [\\
  \ \ \ \{ side: "home",\\
  \ \ \ \ \ result\_bucket: "win",\\
  \ \ \ \ \ goals\_for: 3, ...\}, ...]\},\\
  \ "2009/2010": \{fixtures: [...]\}\}}\hspace{6pt}%
  \panelnote{three query-bearing layers: sparse \emph{dynamic season
  keys}, each holding a \emph{nested fixtures array}; some team
  documents lack the map entirely}%
};
\node[panel={green!11}{black!45}] (expose) at ($(stored.south)+(0,-20)$) {%
  \textbf{\strut Exposed fixture rows}
  \nodepart{two}
  \panelcode{\{ team: "Aberdeen",\\
  \ \ seasons.k: "2008/2009",\\
  \ \ seasons.v.fixtures:\\
  \ \ \ \{ side: "home",\\
  \ \ \ \ \ result\_bucket: "win",\\
  \ \ \ \ \ goals\_for: 3, ... \} \}}\hspace{6pt}%
  \panelnote{one row per \emph{match}: the absent map defaults to
  \{\,\}, season keys become data, then a second unwind at the
  nested-array grain expands each season's fixtures}%
};
\node[panel={orange!14}{black!45}] (grouped) at ($(expose.south)+(0,-20)$) {%
  \textbf{\strut Answer rows}
  \nodepart{two}
  \panelcode{\{ team: "Aberdeen",\\
  \ \ season: "2008/2009",\\
  \ \ played: 38, home\_wins: 13,\\
  \ \ away\_losses: 6,\\
  \ \ goal\_difference: 17 \}\ \ ...}\hspace{6pt}%
  \panelnote{conditional accumulators split home wins from away losses
  per (team, season); thresholds, an eight-key sort, and the top-10
  contract follow}%
};
\draw[flow] ([xshift=-88pt]stored.south) -- ([xshift=-88pt]expose.north);
\node[ops] at ($([xshift=-84pt]stored.south)!0.5!([xshift=-84pt]expose.north)$)
  {{\sffamily\itshape stages 1--3:}\ \$ifNull $\cdot$ \$objectToArray $\cdot$ \$unwind $\cdot$ \$unwind};
\draw[flow] ([xshift=-88pt]expose.south) -- ([xshift=-88pt]grouped.north);
\node[ops] at ($([xshift=-84pt]expose.south)!0.5!([xshift=-84pt]grouped.north)$)
  {{\sffamily\itshape stages 4--9:}\ \$group\,(4$\times$\$cond) $\cdot$ \$addFields\\
   \phantom{{\sffamily\itshape stages 4--9:}\ }$\cdot$ \$match $\cdot$ \$sort $\cdot$ \$limit};
\node[strip, fill=red!4, draw=red!35, anchor=north] (fail) at ($(grouped.south)+(0,-6)$) {%
  \textbf{\textcolor{red!55!black}{Schema-first failures}}\hspace{6pt}a
  fixed-field lookup on \texttt{matches\_by\_season."2008/2009"} cannot
  enumerate unseen season keys, and unwinding at the wrong grain (season
  keys without the nested fixtures, or fixtures pooled across seasons)
  silently corrupts every count.%
};
\end{tikzpicture}%
\caption{A release TEND task (\texttt{european\_football\_2}, nine-stage
gold pipeline). The requested grouping dimension exists only as sparse
dynamic object keys whose values hold nested fixture arrays, so the
pipeline must expose structure as data at two different grains before
conditional aggregation, thresholding, and an ordered top-10 contract.}
\label{fig:tend-example}
\end{figure}
\subsection{Annotation Pipeline}
\label{sec:dataworld_construction}

Seven database-trained annotators proficient in MQL carry out the annotation. The pipeline (Figure~\ref{fig:tend}) is design-first, authoring every MongoDB world before any source data is consulted for population, in five steps:

\textbf{1) Expert DataWorld design.} We author a separate MongoDB construction specification for every database from its domain analytic needs, deciding root entities, collection boundaries, embedded relationships, nesting, dynamic-key conventions, optional paths, sparse fields, and polymorphic shapes by hand. No fixed table-to-collection rule is applied, and no structural decision is inherited from the relational layout.

\textbf{2) Materialization from source evidence.} The expert-designed worlds are then populated with real values drawn from the 11 BIRD mini-dev databases, whose contents and workload evidence ground each domain; no data synthesis is needed, and BIRD SQL is treated only as evidence of analytic intent, never as MQL templates or execution oracles.

\textbf{3) LLM-assisted NL--MQL drafting and human refinement.} LLMs then mine candidate analytic intents from the source evidence and draft initial MQL programs at scale; no candidate is accepted as gold. We manually refine the intent, revise the pipeline against witness execution, and write a canonical English request aligned to the DataWorld.

\textbf{4) Execution-based verification.} Every gold MQL is executed on frozen MongoDB witness data and its result inspected; the query or wording is revised and re-checked until intent, program, and execution result are mutually consistent.

\textbf{5) Benchmark audit, repair, and release quality control.} After construction, multiple full-benchmark audit-and-repair passes (deterministic checks plus LLM-assisted review under expert adjudication) targeted NL--MQL alignment defects, e.g., answers whose stored form or ordering is not decidable from the question, re-verifying every repaired task by execution. Deterministic validation then enforces the public record format, parses every pipeline, rejects duplicate MQL signatures, controls skeleton-family concentration, and verifies exact execution on the witness data; the release contains 1,210 validated tasks, each with one gold MQL program and one canonical English request.

\newlength{\statbarlen}\setlength{\statbarlen}{78pt}
\definecolor{statbarfill}{RGB}{86,118,176}
\newcommand{\statbar}[1]{{%
  \rlap{\textcolor{black!9}{\rule[0.25ex]{\statbarlen}{4.4pt}}}%
  \textcolor{statbarfill}{\rule[0.25ex]{#1\statbarlen}{4.4pt}}}}
\newcommand{\statpanel}[1]{\multicolumn{4}{@{}l}{\cellcolor{black!6}\textsc{#1}}}

\begin{table}[t]
    \centering
    \caption{Release-scale statistics of the TEND benchmark.}
    \label{tab:tend-overview}
    \footnotesize
    \renewcommand{\arraystretch}{1.06}
    \setlength{\tabcolsep}{3pt}
    \begin{tabularx}{\columnwidth}{@{}>{\raggedright\arraybackslash}Xr@{\hspace{10pt}}>{\raggedright\arraybackslash}Xr@{}}
    \toprule
    \statpanel{Scale and validation} \\
    Databases & 11 & NL--MQL tasks & 1,210 \\
    Tasks per database & 110 & Collections & 32 \\
    Witness documents & 269,177 & Exact execution & \multicolumn{1}{r}{1,210\,/\,1,210} \\
    \midrule
    \statpanel{Gold-program complexity and diversity} \\
    Stages (med.\,/\,max) & 7\,/\,13 & Operators (med.) & 12 \\
    Unique MQL signatures & 1,210 & Skeleton families & 1,098 \\
    Largest skeleton fam. & 6\,(0.5\%) & Skeleton entropy & 0.99 \\
    \midrule
    \end{tabularx}\par\nointerlineskip
    \begin{tabularx}{\columnwidth}{@{}>{\raggedright\arraybackslash}Xrr@{\hspace{7pt}}l@{}}
    \statpanel{Native-operator usage (records)} \\
    Array operators & 1,170 & 96.7\% & \statbar{0.967} \\
    Nested dotted paths & 1,164 & 96.2\% & \statbar{0.962} \\
    Dynamic-key operators & 1,096 & 90.6\% & \statbar{0.906} \\
    Grouping accumulators & 652 & 53.9\% & \statbar{0.539} \\
    \midrule
    \statpanel{Claim-axis slices (records)} \\
    \multicolumn{4}{@{}l}{\emph{Schema-flex category}} \\
    \quad Dynamic-key map & 1,013 & 83.7\% & \statbar{0.837} \\
    \quad Nested event stream & 130 & 10.7\% & \statbar{0.107} \\
    \quad Polymorphic shape & 36 & 3.0\% & \statbar{0.030} \\
    \quad Missing-vs-present & 18 & 1.5\% & \statbar{0.015} \\
    \quad Attribute bag & 13 & 1.1\% & \statbar{0.011} \\
    \multicolumn{4}{@{}l}{\emph{Anti-SQL-transfer level}} \\
    \quad Strong & 1,138 & 94.0\% & \statbar{0.940} \\
    \quad Medium & 71 & 5.9\% & \statbar{0.059} \\
    \quad Weak & 1 & 0.1\% & \statbar{0.001} \\
    \bottomrule
    \end{tabularx}
\end{table}

\subsection{Dataset Statistics and Characteristics}
\label{sec:tend_statistics}

\begin{table*}[t]
\centering
\caption{Comparison of TEND with existing benchmarks.}
\label{tab:benchmark-comparison}
\renewcommand{\arraystretch}{1.18}
\setlength{\tabcolsep}{5pt}
\resizebox{\textwidth}{!}{%
\begin{threeparttable}
\begin{tabular}{l c c c c c c c}
\toprule
\textbf{Dataset} &
\textbf{Target} &
\textbf{Scale} &
\textbf{\makecell[c]{Native\\DW\tnote{a}}} &
\textbf{\makecell[c]{Schema\\Flex.\tnote{b}}} &
\textbf{\makecell[c]{Nested\\Fields}} &
\textbf{\makecell[c]{Program Diversity\\Control}} &
\textbf{\makecell[c]{Construction\\Leader\tnote{c}}} \\
\midrule
WikiSQL \cite{zhong2017seq2sql} & SQL & 80k+ NLQs & \xmark & \xmark & \xmark & \xmark & Program \\
Spider \cite{yu-etal-2018-spider} & SQL & 10,181 NLQs & \xmark & \xmark & \xmark & SQL complexity & Human \\
BIRD \cite{NEURIPS2023_83fc8fab} & SQL & 12,751 NLQs & \xmark & \xmark & \xmark & SQL complexity & Human \\
KaggleDBQA \cite{lee-etal-2021-kaggledbqa} & SQL & 272 NLQs & \xmark & \xmark & \xmark & \xmark & Human \\
Spider 2.0 \cite{lei2025spider} & SQL & 632 workflows & \xmark & \xmark & \xmark & Workflow complexity & Human \\
DocSpider \cite{ozer2025docspider} & MQL & 4,663 NLQs & \xmark & \xmark & \xmark & Inherited from Spider & LLM+Program \\
SM3-Text-to-Query \cite{sivasubramaniam2024sm3} & \makecell[c]{SQL/MQL\\Cypher/SPARQL} & \makecell[c]{10K NLQs\\40K pairs} & \xmark & Limited & Partial & Template augmented & Human+Program \\
\midrule
\rowcolor{blue!5}
\textbf{TEND} & \textbf{MQL} & \textbf{1,210 NLQs} & \cmarkbold & \cmarkbold & \cmarkbold & \textbf{1,210 unique MQL signatures} & \textbf{Human+LLM} \\
\bottomrule
\end{tabular}
\begin{tablenotes}
\footnotesize
\item[a] \textbf{Native DW:} the database worlds are MongoDB-native DataWorlds authored for document semantics, rather than relational mirrors produced by one-collection-per-table migration.
\item[b] \textbf{Schema Flex.:} query-bearing schema flexibility: dynamic keys, optional fields, and polymorphic shapes participate in gold programs; ``Limited'' denotes template-level variation only.
\item[c] \textbf{Construction Leader:} Human: expert-authored; Program: rule-based generation; Human+Program: templates with programmatic augmentation; LLM+Program: LLM-translated gold gated by execution checks; Human+LLM: human-led with AI assistance and expert validation.
\end{tablenotes}
\end{threeparttable}%
}
\end{table*}

Table~\ref{tab:tend-overview} summarizes TEND at release scale: 1,210 execution-distinct tasks balanced over 11 databases (110 each), each with one canonical English request, and 269,177 witness documents across 32 collections; every gold program is an aggregation pipeline with a unique MQL signature that passes exact execution on the frozen witness data. The median pipeline has seven stages and twelve distinct operators, requiring multi-stage document transformation rather than shallow filtering, and MongoDB-native stress is pervasive: 90.6\% of records use dynamic-key operators, 96.7\% array operators, and 96.2\% nested dotted paths.

Program diversity is controlled at the template grain as well: a \emph{skeleton family} groups gold programs that become identical after abstracting field names and literals; the largest family holds six of the 1,210 records and the normalized skeleton entropy is 0.99 (Table~\ref{tab:tend-overview}), so TEND cannot be solved by memorizing a small set of pipeline templates.

\textbf{Claim axes.} Every record carries two designed labels that the evaluator uses for sliced reporting, with release distributions in Table~\ref{tab:tend-overview}. The \emph{schema-flex category} names the document-native structure the task exercises: dynamic-key maps (the dominant category), nested event streams, polymorphic shapes, missing-versus-present semantics, and attribute bags. The \emph{anti-SQL-transfer level} grades, from gold-program operator evidence, how strongly the task resists relational transfer: a record is \emph{strong} when its pipeline combines a strongly document-native operator (e.g., \texttt{\$objectToArray} or \texttt{\$switch}) with further native structure, \emph{medium} when it uses native expressions without such combination, and \emph{weak} when it reduces to relational-style flatten-and-group. These labels are evaluation metadata only and are never visible to any solver or baseline.

Table~\ref{tab:benchmark-comparison} summarizes these distinctions: \textbf{MongoDB-native DataWorld design} (collections, nesting, and query semantics authored for MongoDB rather than inherited from relational catalogs or one-collection-per-table migrations), \textbf{query-bearing schema flexibility} (dynamic keys, optional fields, arrays, and nested paths participate in gold MQL programs rather than appearing as passive irregularities), and \textbf{program-diversity control} (1,210 unique MQL signatures, ruling out template-like generation). The construction is human-led with LLM assistance, followed by expert validation and frozen MongoDB execution.

\begin{figure*}[t]
    \centering
    \includegraphics[width=0.8\textwidth]{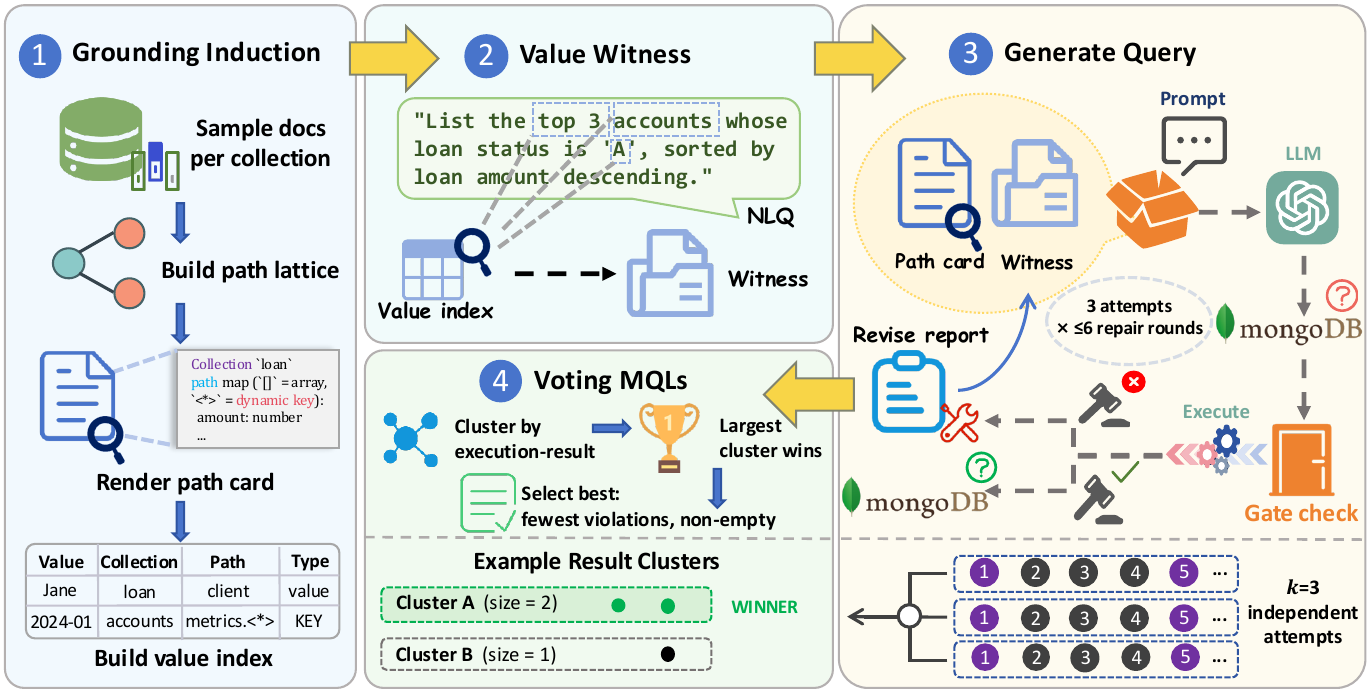}
    \caption{SAG inference under Schema-as-Data Grounding. SAG first induces a path and value grounding index from stored MongoDB witnesses, then uses grounded value witnesses, gated generation, execution-based repair, and result-cluster voting to select a deterministic MQL prediction.}
    \label{fig:sag}
\end{figure*}

\subsection{Evaluation Metrics}
\label{sec:tend_eval_metrics}

TEND uses an execution-grounded evaluator rather than relying only on textual similarity to the gold program. For each benchmark record, let $q_p$ denote the predicted MongoDB program, $q_g$ the gold MQL program, and $D$ the frozen MongoDB witness database. The evaluator parses $q_p$, checks it against the disabled-operator list, and executes both $q_p$ and $q_g$ with the same normalized executor $\operatorname{NormExec}(\cdot,D)$, which canonicalizes BSON/JSON values and numeric types while preserving semantically meaningful distinctions such as null versus missing fields and retained \texttt{\_id} values. Parse failures, forbidden operators, execution errors, missing predictions, and typed solver failures are all kept as zero-score per-record rows, so reported averages preserve the full denominator. The metric suite consists of one headline metric and three companions, all derived from the same execution results so they require no extra runs.

\paragraph{EXC (headline): bounded column-tolerant execution accuracy} EXC follows the execution-equivalence intuition of database benchmarks, where top-level output column names are not by themselves semantic. A prediction receives $\mathrm{EXC}=1$ iff it (i) parses, (ii) contains no forbidden operator, (iii) returns the same number of rows as the gold result, and (iv) aligns every gold row with a predicted row whose top-level value multiset contains all gold values with at most $\beta=2$ surplus top-level values. The surplus bound accounts for benign MongoDB aggregation artifacts, such as a leftover \texttt{\_id} group key or one helper/sort key, while still rejecting projection-free document dumps. Row order is enforced when the gold pipeline contains order-sensitive root stages (\texttt{\$sort}, \texttt{\$limit}, \texttt{\$skip}, \texttt{\$setWindowFields}); otherwise rows are matched as multisets. Nested object keys, array order, value identity, row count, and missing values remain strict, so field-name tolerance cannot hide an incorrect answer. EXC is a per-record binary indicator averaged uniformly over the benchmark.
\paragraph{EXF\textsubscript{1}: graded result-set $F_1$} Because EXC is binary, it cannot distinguish a near-miss from a completely wrong answer. EXF\textsubscript{1} is the order-insensitive multiset $F_1$ between predicted and gold result rows under the same row identity as EXC (top-level column names cosmetic, nested keys and values significant) but with no surplus tolerance. It grades how close a wrong answer is to the gold result set while collapsing to zero precision on projection-free dumps, so it cannot be gamed by returning extra rows.
\paragraph{Execution-outcome decomposition} Every scored record falls into exactly one mutually exclusive bucket: \emph{correct} (the EXC successes), \emph{no submission}, \emph{invalid} (parse failure or forbidden operator), \emph{execution error}, \emph{empty result}, or \emph{wrong rows} (executed but the row multiset or its values disagree, reported as structural near misses, i.e., order-only, row-subset, and row-superset, and as value mismatches). The bucket fractions sum to one, turning the headline gap $1-\mathrm{EXC}$ into an interpretable loss budget that reports \emph{where} along the schema-less reasoning claim systems fail.
\paragraph{Claim-axis slices} EXC and EXF\textsubscript{1} are sliced along the benchmark's claim axes of schema-flex category, anti-SQL-transfer level, and database domain (Section~\ref{sec:tend_statistics}), with each slice reported alongside its record count, so an aggregate score cannot hide collapse on a rare native structure such as the 13-record attribute-bag slice.

For the ablation study, where neighboring ladder arms can differ by only a handful of records, we additionally use an exact paired McNemar test of each arm's EXC against SAG, so claimed mechanism contributions are separated from noise. We deliberately do not report query-form or naming diagnostics (e.g., exact program match or field-naming agreement): program-form match is near zero for any independently generated query and output-column naming is underdetermined by the request, so neither discriminates between systems.

\section{SAG: Schema-as-Data Grounding}
\label{sec:sag}

SAG addresses Text-to-NoSQL as evidence-constrained program synthesis rather than one-shot query generation. Given a natural-language query and read-only access to a MongoDB database, the goal is to produce a deterministic aggregation pipeline that answers the query while respecting schema-less document structure. This setting differs from NL2SQL inference because the operative schema is not a fixed catalog: fields may be sparse, nested, repeated, dynamically keyed, or visible only after aggregation stages reshape the document stream. SAG therefore treats stored documents as schema evidence. Database contents induce the admissible path space, value evidence, and repair signals that constrain decoding and validation.

\subsection{Overview}
\label{sec:sag_overview}

Figure~\ref{fig:sag} shows the end-to-end solver. SAG is organized around a per-database grounding index and a bounded per-query decode--gate--repair procedure. The index is built from bounded witness sampling over the read-only database. It records the collection set, the induced path lattice, dynamic-key maps collapsed to \texttt{<*>}, top-level fields, value witnesses, and compact card text that is provided to the decoder. During inference, the model receives only the NLQ, this induced card, and optional value-witness lines; gold programs and benchmark annotations are excluded from the input.

The design has four principles: schema-less structure is treated as data-induced perception, with every admissible field path represented in the induced lattice and dynamic maps exposed through collapsed paths and example keys; value grounding anchors hard NLQ literals to exact stored forms, including dynamic-map keys; deterministic gates check path and value admissibility before execution; and execution feedback serves as a repair gradient, with repaired samples clustered by result equivalence so the final answer is selected in result space rather than by surface form. Unlike ReAct, SAG grounds deterministically before generation and reserves the loop for bounded correction of concrete violations.

\subsection{Problem Formulation and Grounding Objective}
\label{sec:sag_formulation}

Let a MongoDB database be $D=\{C_1,\ldots,C_m\}$, where each collection $C_i$ is a multiset of JSON-like documents rather than a fixed relation. A user utterance is denoted by $x$, and the target output is a read-only aggregation program $p=(c,\mathbf{s})$ consisting of a base collection $c\in D$ and an ordered stage sequence $\mathbf{s}=(s_1,\ldots,s_T)$. Before decoding, SAG induces a grounding index
\begin{equation}
\mathcal{I}_D=(\mathcal{C},\mathcal{L},\mathcal{P},\mathcal{M},\mathcal{V},\mathrm{card}),
\label{eq:sag-index}
\end{equation}
where $\mathcal{C}$ is the closed collection set, $\mathcal{L}$ is the induced path lattice, $\mathcal{P}$ contains valid path prefixes, $\mathcal{M}$ records dynamic-key maps and example keys, $\mathcal{V}$ is the indexed set of witnessed stored values and dynamic keys, and $\mathrm{card}$ is the rendered path card used as the decoder's hypothesis space. A candidate program induces read claims over collections, paths, values, limits, and joins:
\begin{equation}
\Gamma(p)=\Gamma_{\mathrm{coll}}(p)\cup\Gamma_{\mathrm{path}}(p)\cup\Gamma_{\mathrm{value}}(p)\cup\Gamma_{\mathrm{limit}}(p)\cup\Gamma_{\mathrm{join}}(p).
\label{eq:sag-claims}
\end{equation}
The alignment gate implements an admissibility predicate
\begin{equation}
A(p;\mathcal{I}_D,x)=A_{\mathrm{path}}(p;\mathcal{I}_D)\wedge A_{\mathrm{value}}(p;\mathcal{I}_D,x)\wedge A_{\mathrm{limit}}(p,x),
\label{eq:sag-admissible}
\end{equation}
where $A_{\mathrm{path}}$ checks the selected collection, field references, computed intermediates, dynamic-key accesses, and data-witnessed \texttt{\$lookup} edges; $A_{\mathrm{value}}$ requires hard witnessed literals to appear with exact stored forms on witnessed paths; and $A_{\mathrm{limit}}$ enforces explicit top/first/limit requests in the NLQ. Each predicate has a constructive counterpart $V_{\bullet}$ that returns the set of concrete violations it detects, so that $A_{\bullet}$ holds exactly when $V_{\bullet}=\emptyset$; these violation sets are what the repair loop consumes (Algorithm~\ref{alg:sag}). Probe failures are fail-open so that the gate does not reject gold-equivalent programs merely because an auxiliary witness probe is unavailable.

SAG therefore searches over read-only deterministic aggregation programs under admissibility and execution feedback:
\begin{equation}
\begin{aligned}
p^{*}=\arg\max_{p\in\Pi_{\mathrm{ro}}(D)}
\Big[&
S_{\theta}(p\mid x,\mathrm{card},W_x)\\
&+\lambda_{\mathrm{gate}}\mathbb{I}\{A(p;\mathcal{I}_D,x)\}\\
&+\lambda_{\mathrm{exec}}\mathbb{I}\{\operatorname{Exec}_D(p)\ \mathrm{nonempty}\}
\Big],
\end{aligned}
\label{eq:sag-objective}
\end{equation}
where $\Pi_{\mathrm{ro}}(D)$ is the space of read-only aggregation programs over $D$, $W_x$ denotes the value witnesses extracted by matching literals in $x$ against $\mathcal{V}$, and $E_x\subseteq W_x$ is the enforceable subset of hard witnesses (those with few enough locations to gate on). SAG realizes Eq.~\ref{eq:sag-objective} procedurally: the LLM proposes structured candidates, gates return repair messages for inadmissible candidates, execution exposes execution errors or empty outputs, and the selected candidate minimizes violations and empty results across bounded repair rounds.

\begin{algorithm}[t]
    \small
    \SetKwInOut{Input}{Input}
    \SetKwInOut{Output}{Output}
    \Input{User utterance $x$, read-only database $D$, policy $(k,R)$}
    \Output{Aggregation program $p$ or failure diagnostic}
    $\mathcal{I}_D\leftarrow\textsc{BuildGroundingIndex}(D)$\;
    $(W_x,E_x)\leftarrow\textsc{WitnessLiterals}(x,\mathcal{I}_D)$\;
    $\mathcal{O}\leftarrow\emptyset$\;
    \For{$i\leftarrow1$ \KwTo $k$}{
        $M\leftarrow[\textsc{SystemPrompt}(\mathcal{I}_D),\textsc{Question}(x,W_x)]$\;
        $\mathcal{B}\leftarrow\emptyset$\;
        \For{$r\leftarrow1$ \KwTo $R$}{
            $T\leftarrow\bot$\;
            $p\leftarrow\textsc{DecodeStructured}(M,\mathcal{C})$\;
            $F\leftarrow V_{\mathrm{path}}(p,\mathcal{I}_D)\cup V_{\mathrm{value}}(p,E_x)\cup V_{\mathrm{limit}}(p,x)$\;
            \If{$F=\emptyset$}{
                $T\leftarrow\textsc{ExecuteReadOnly}(p,D)$\;
                \If{$T$ is empty}{$F\leftarrow\{\textsc{BisectEmpty}(p,D,\mathcal{I}_D)\}$}
                \ElseIf{$T$ leaks synthetic \texttt{\_id}}{$F\leftarrow\{\textsc{SuppressId}\}$}
            }
            $\mathcal{B}\leftarrow\mathcal{B}\cup\{(p,F,T)\}$\;
            \If{$F=\emptyset$}{break}
            $M\leftarrow M\cup[\textsc{Candidate}(p),\textsc{RepairFeedback}(F)]$\;
        }
        $\mathcal{O}\leftarrow\mathcal{O}\cup\{\textsc{BestByViolations}(\mathcal{B})\}$\;
    }
    \If{$\mathcal{O}=\emptyset$}{\Return{\textsc{Fail}(no valid candidate)}}
    \Return{\textsc{LargestResultCluster}$(\mathcal{O})$}
    \caption{SAG inference under Schema-as-Data Grounding.}
    \label{alg:sag}
\end{algorithm}

Algorithm~\ref{alg:sag} summarizes inference. Unlike ordinary agent prompting, structure discovery is not delegated to an in-loop exploratory agent: the complete data-induced card is provided before decoding, and the loop only corrects concrete violations (nonexistent paths, inadmissible joins, wrong literal bindings, missing limits, execution errors, empty results, and leaked identifiers).

\subsection{Grounding Index and Path Card}
\label{sec:sag_index}

The grounding index is the schema-as-data representation constructed for each database. SAG samples documents from each collection and recursively walks objects and arrays up to a bounded depth. Each observed node becomes a lattice entry with a dominant type, count, and, for object nodes, a static-document or dynamic-map classification. Dynamic maps are detected from key proliferation and data-like keys, then rendered with the placeholder \texttt{<*>}; this prevents concrete keys such as dates, codes, and map entries from exhausting the context while still revealing the access pattern.

From the lattice, SAG derives three inference artifacts. The valid-prefix set supports path membership checks; the top-field set distinguishes physical input fields from computed intermediates introduced later in an aggregation pipeline; and the path card renders the entire induced lattice as decoder context. The decoder instruction specifies that paths absent from the induced card should be treated as nonexistent, that related data are usually embedded rather than joined, and that stored values must be copied verbatim unless the question explicitly defines a label mapping.

The card is intentionally stronger than a name-level schema summary: induced from the same read-only database the generated query executes on, it includes nested array paths, exposes sparse and presence-wrapped fields when observed, and represents dynamic-key structures compactly.

\subsection{Value Witnesses and Alignment Gates}
\label{sec:sag_evidence}

SAG complements path grounding with value-witness anchoring: quoted strings, dates, all-caps codes, and selected title-case phrases from the NLQ are normalized and looked up in the value index. A witness records where the term occurs, whether it is a stored value or a dynamic key, and the exact stored form; hard witnesses with few locations are enforceable, while soft witnesses inform the model but do not block candidates.

The deterministic gate has two main sides. $A_{\mathrm{path}}$ checks that the selected collection exists, every outer-pipeline field reference resolves to an induced path or valid computed intermediate, and each \texttt{\$lookup} targets a real collection through a data-witnessed join edge. If a join edge cannot be probed safely, the gate fails open, but nonexistent collections and missing foreign fields remain violations. $A_{\mathrm{value}}$ checks comparison-position strings against hard witnesses so that a model cannot observe a literal in one path and silently filter a different path or casing.

A third contract enforces explicit row limits: a top-$N$, first-$N$, or at-most-$N$ request must carry the corresponding \texttt{\$limit}. Together these checks turn common failure modes into model-facing feedback: wrong collection names, hallucinated paths, unnecessary joins over embedded data, misbound constants, and missing limit semantics.

\subsection{Execution Repair and Result Consistency}
\label{sec:sag_repair}

After a candidate passes the alignment gate, SAG scans for disabled operators and executes the pipeline against the read-only world. Execution is not used as a gold-answer oracle; it is a validity and repair signal. Execution errors are returned as concise feedback. Empty results trigger prefix bisection: SAG executes successive pipeline prefixes with \texttt{\$count} to locate the first stage that collapses the stream to zero rows or errors, and, when possible, reports actual distinct values at filtered paths. This converts an uninformative ``0 rows'' outcome into a localized, data-grounded repair gradient.

SAG also enforces an output contract for internal identifiers: a pipeline that leaks synthetic \texttt{\_id} values is repaired unless the question asks for them. Repair feedback includes the previous candidate and its concrete violations and requests a corrected structured candidate; within an attempt, SAG retains the candidate with the fewest violations, preferring non-empty executions and later rounds.

SAG uses $k=3$ attempts in its full configuration: final executions are clustered by order-insensitive result equivalence, the largest cluster supplies the prediction, and ties prefer fewer violations and non-empty results. Selection happens in result space because MongoDB programs with different surface forms can compute the same answer while plausible pipelines diverge after an unwind, projection, or group. If all attempts fail, SAG reports a failure diagnostic rather than an unsupported query.

\section{Experiments and Analysis}
\label{sec:exp_analysis}
This section evaluates SAG on the TEND release, compares it with controlled information-channel baselines, measures SAG mechanism ablations, and reports EXC together with diagnostic metrics. The results show that TEND is a difficult benchmark even for a strong reasoning model: the best system remains below 36\% mean EXC, while SAG obtains the highest mean accuracy, the highest graded result quality, and the cleanest executable behavior under the same model and evaluator.

\subsection{Experimental Setup}
\label{sec:exp_setup}
\subsubsection{Dataset}
Experiments use the public \emph{tend-native-mongodb-v1} release (1,210 execution-distinct tasks over 11 databases; Section~\ref{sec:tend_statistics}). All systems are evaluated on the identical task set against the same frozen MongoDB witness data, with missing predictions, execution errors, and typed failures retained in the denominator as zero-score rows.

\subsubsection{Models}
All compared systems share the same backbone LLM, decoding configuration, and evaluator; they differ only in their \emph{information channel}, that is, in what each method is allowed to observe about the schema-less MongoDB database, and in whether that channel is static (one-shot) or interactive (exploration or repair). No method receives the gold MQL, the canonical-form set, or any release annotation. The compared methods are:

\begin{itemize}
    \item \textbf{NLQ-only Direct.} A channel-purity floor and contamination probe: only the natural-language question, with no schema, samples, or database access; the pipeline is emitted in one shot. Non-trivial accuracy here would indicate memorization rather than grounding.
    \item \textbf{Schema Direct.} The question plus a name-level structural summary of the database (collection names and document field names, no stored values or samples), isolating how far name-level structure alone carries.
    \item \textbf{Sampled-doc Direct.} The question plus three raw documents per collection, from which collections, field paths, and value formats must be inferred in one shot.
    \item \textbf{Data-rich Direct.} Identical except for a larger sample budget (eight documents per collection), separating sample size from mechanism.
    \item \textbf{SQL Pivot.} The same channel as Sampled-doc Direct, but the model first organizes the intent as a self-generated relational SQL sketch and then translates it into MQL, with no gold SQL, relational source schema, external converter, or execution feedback. This arm measures whether a relational intermediate helps or hurts document-native reasoning (Section~\ref{sec:exp_analysis}).
    \item \textbf{ReAct.} An agentic exploration baseline and the apples-to-apples counterpart of SAG: given the same collection-name list SAG receives, it self-paces a bounded thought--action--observation loop of read-only actions with \emph{raw} first-five-row observations. Budget exhaustion without a submitted pipeline is a typed failure, never converted into a forced answer. Unlike SAG it receives no induced path card, value witnesses, alignment gates, or repair feedback.
    \item \textbf{SAG (ours).} The proposed solver in its full configuration (Section~\ref{sec:sag}): induced path-lattice card and value index from a bounded witness sample, value witnesses for question literals, the joint path-and-value admissibility gate with the limit contract, at most six repair rounds with prefix-bisection feedback on empty results, and three-sample result-space consistency. All probes are read-only; execution is repair evidence only, never a correctness oracle.
\end{itemize}

All non-agentic baselines see only sanitized public record fields and emit machine-checkable disclosure fields stating exactly which channel they observed; ReAct's raw-row visibility is likewise declared in its disclosure.

\subsubsection{Evaluation Metrics}
\label{sec:metric}
The headline execution metric is EXC, the bounded column-tolerant execution accuracy defined in Section~\ref{sec:tend_eval_metrics}. We additionally report the graded result-set $F_1$ (EXF\textsubscript{1}) and a mutually exclusive execution-outcome decomposition. In the compact tables, NoSub denotes missing or typed no-submission failures, Exec denotes MongoDB execution errors, Empty denotes executable pipelines that return no rows, Struct. aggregates order-only, row-subset, and row-superset near misses, and Value denotes value-mismatch failures. Invalid candidates are retained in the denominator; they are 0.0\% for all headline rows in this snapshot and are therefore omitted from the compact tables. The bucket fractions sum to one with EXC, and ablation significance statements use an exact paired McNemar test of each arm's EXC against SAG.

\begin{table}[t!]
    \centering
    \caption{Main comparison on TEND. Best accuracy values are bolded.}
    \label{tab:results}
    \setlength{\tabcolsep}{2.2pt}
    \renewcommand{\arraystretch}{0.96}
    \resizebox{\columnwidth}{!}{
        \begin{tabular}{@{}lccccccc@{}}
        \toprule
        & \multicolumn{2}{c}{\textbf{Accuracy}} & \multicolumn{5}{c}{\textbf{Outcome decomposition (\%)}} \\
        \cmidrule(lr){2-3}\cmidrule(l){4-8}
        \textbf{Method} & EXC & EXF\textsubscript{1} & NoSub & Exec & Empty & Struct. & Value \\
        \midrule
        NLQ-only Direct & 0.2 & 0.2 & 8.2 & 1.1 & 89.4 & 0.0 & 1.2 \\
        Schema Direct & 0.2 & 0.4 & 6.3 & 15.3 & 69.3 & 0.1 & 8.8 \\
        Sampled-doc Direct & 28.6 & 29.4 & 2.1 & 4.7 & 7.2 & 1.6 & 55.7 \\
        Data-rich Direct & 28.9 & 29.8 & 4.8 & 3.1 & 3.8 & 1.6 & 57.7 \\
        SQL Pivot & 25.6 & 26.2 & 1.5 & 4.9 & 8.8 & 1.4 & 57.8 \\
        ReAct & 30.8 & 34.6 & 9.0 & 0.2 & 0.3 & 2.7 & 56.9 \\
        \midrule
        \textbf{SAG (Ours)} & \textbf{35.8} & \textbf{40.5} & 0.0 & 0.0 & 0.5 & 2.2 & 61.5 \\
        \bottomrule
      \end{tabular}
    }
\end{table}

\subsubsection{Implementation Details}
All systems use the same backbone LLM, DeepSeek-V4-Flash, a reasoning model accessed through an OpenAI-compatible API with reasoning effort \emph{max}, temperature $0.0$, and no output-token cap (a cap truncates long reasoning traces and artificially inflates no-submission failures); the identical client, decoding configuration, and evaluator make reported differences attributable to the information channel and mechanism alone. SAG runs with $k{=}3$ attempts, at most six repair rounds per attempt, a 400-document witness sample per collection, a 400-entry path card, and a 20-second per-query execution timeout; ReAct receives 50 interaction steps per task. All probing and execution are read-only against the frozen release witness data in a live MongoDB instance; no system writes to the database or observes evaluation output.

\subsection{Performance Comparison}

Table~\ref{tab:results} reports the main comparison. \textbf{The channel-purity floors show that TEND can only be solved by grounding.} NLQ-only Direct and Schema Direct both reach 0.2\% EXC, so the release is not recoverable from language priors (a clean contamination probe) and name-level structure is nearly worthless: 69.3\% of Schema Direct pipelines execute to empty results because collection and field names reveal neither dynamic keys, populated paths, nor stored value forms. Even with grounding the benchmark stays hard. The best system solves 35.8\%, no method exceeds 54.5\% on any database, and for every value-bearing channel the dominant failure bucket is wrong rows, that is, pipelines that parse, execute, and return plausible but incorrect results, the failure regime only execution-result evaluation exposes.

\textbf{More data without mechanism saturates immediately; SAG's gain is mechanism.} Nearly tripling the one-shot sample (Sampled-doc to Data-rich Direct) buys +0.3 points, and 50 steps of interactive exploration (ReAct) buy +1.9 more at the panel's highest no-submission rate (9.0\%, budget exhaustion). SAG reaches 35.8\% EXC and 40.5 EXF\textsubscript{1}, a gain of +5.0 EXC over ReAct and +6.9 over the strongest one-shot channel, while replacing open-ended exploration with bounded repair (at most six rounds, averaging close to one). Its EXF\textsubscript{1}$-$EXC gap (4.7 points, the largest in the panel) shows that even SAG's misses are near-misses in result space rather than structural derailments. The interaction economics compound the accuracy gap: in our development runs the SAG arms averaged 1.1 to 1.2 repair rounds per attempt against 13 to 15 consumed ReAct steps per task, so grounding before decoding is not only more accurate but roughly an order of magnitude cheaper in model interactions.

\textbf{SAG eliminates the mechanical failure surface.} It is the only system with zero no-submission, invalid, or execution-error rows; its residual loss is concentrated in the semantic wrong-rows bucket where the real headroom lies. At the database grain (all systems run one fixed configuration across the eleven databases, with no per-database tuning), SAG is best or tied-best on seven of eleven, with its largest leads where document-native structure dominates: 50.9\% versus 36.4\% for the best baseline on \texttt{toxicology}, 50.0\% versus 41.8\% on \texttt{european\_football\_2}, and 33.6\% versus 20.9\% on \texttt{superhero}, while matching the best baseline on \texttt{financial} (54.5\%). Per-database EXC ranges from 17.3\% (\texttt{formula\_1}) to 54.5\%, ReAct retains an edge on \texttt{formula\_1} and \texttt{thrombosis\_prediction}, and no method dominates every database, the intended behavior of a benchmark whose DataWorlds stress different combinations of dynamic keys, embedded evidence, sparse paths, and aggregation shape.

\subsection{Where the Difficulty Comes From}

\begin{table}[t!]
    \centering
    \caption{EXC (\%) by claim-axis slice.}
    \label{tab:slices}
    \setlength{\tabcolsep}{2.2pt}
    \renewcommand{\arraystretch}{0.96}
    \begin{threeparttable}
    \resizebox{\columnwidth}{!}{
        \begin{tabular}{@{}lccccccc@{}}
        \toprule
        & \multicolumn{5}{c}{\textbf{Schema-flex category}} & \multicolumn{2}{c}{\textbf{Anti-SQL}} \\
        \cmidrule(lr){2-6}\cmidrule(l){7-8}
        \textbf{Method} & \makecell{DynKey\\(1013)} & \makecell{Event\\(130)} & \makecell{Poly\\(36)} & \makecell{Missing\\(18)} & \makecell{AttrBag\\(13)} & \makecell{Med.\\(71)} & \makecell{Strong\\(1138)} \\
        \midrule
        Data-rich Direct & 29.1 & 27.7 & 25.0 & 44.4 & 15.4 & 23.9 & 29.2 \\
        SQL Pivot & 26.0 & 22.3 & 33.3 & 33.3 & 0.0 & 14.1 & 26.3 \\
        ReAct & 28.5 & \textbf{45.4} & \textbf{41.7} & 44.4 & 15.4 & \textbf{36.6} & 30.4 \\
        \midrule
        \textbf{SAG (Ours)} & \textbf{35.5} & 38.5 & 22.2 & 44.4 & \textbf{53.8} & 29.6 & \textbf{36.1} \\
        \bottomrule
      \end{tabular}
    }
    \begin{tablenotes}\footnotesize
    \item DynKey = dynamic-key maps, Event = nested event streams, Poly = polymorphic shapes, Missing = missing-vs-present, AttrBag = attribute bags; record counts in parentheses. The single-record \emph{weak} transfer slice is omitted.
    \end{tablenotes}
    \end{threeparttable}
\end{table}

Table~\ref{tab:slices} resolves the gap along the claim axes. \textbf{SAG's margin concentrates exactly where schema-less structure carries the query}: +6.4 points over the best baseline on the dominant dynamic-key slice (1,013 of 1,210 records), 53.8\% versus 15.4\% on attribute bags, and +5.7 on the strong anti-SQL-transfer slice (94\% of the release). These are the slices where a name-level catalog is least informative, because the query-bearing structure lives in object keys, array contents, embedded records, and stored values. ReAct stays competitive on the two smallest structural families, nested event streams ($n{=}130$) and polymorphic shapes ($n{=}36$), where iteratively probing a handful of co-existing document variants can locally substitute for global lattice induction.

The headline is not a metric artifact. All 1,210 gold pipelines self-score $\mathrm{EXC}{=}1$, an independent $\beta$-sweep selects the frozen surplus bound $\beta{=}2$, and counterfactual witness-injection flips only 3 of 201 SAG-correct verdicts, so the reported gains are not artifacts of empty outputs, helper-column tolerance, or fragile database instances.

\subsection{Ablation Study}

We ablate SAG through a cumulative mechanism ladder whose settings are named by the strongest capability available in each arm. \textbf{Path Card} gives the decoder the induced path card but performs a single ungated generation. \textbf{Gated Repair} adds path admissibility checking and execution-grounded repair with plain empty-result feedback. \textbf{Value/Bisection Repair} further adds value witnesses, value alignment, the limit contract, and prefix-bisection feedback, but keeps a single final attempt. \textbf{SAG} is the complete proposed solver, adding three-way result-consistency clustering. Table~\ref{tab:ablation_study} uses the same execution-result vocabulary as the main comparison table, so each ablation result is read as both an accuracy change and a failure-mode redistribution. Because adjacent ladder arms can differ by only a handful of records, significance statements use an exact paired McNemar test of each arm's per-record EXC against SAG rather than raw deltas alone.

\begin{table}[t!]
    \centering
    \caption{Ablation of SAG.}
    \label{tab:ablation_study}
    \setlength{\tabcolsep}{2.0pt}
    \renewcommand{\arraystretch}{0.96}
    \begin{threeparttable}
    \resizebox{\columnwidth}{!}{
        \begin{tabular}{@{}lccccccc@{}}
        \toprule
        & \multicolumn{2}{c}{\textbf{Accuracy}} & \multicolumn{4}{c}{\textbf{Failures (\%)}} & \textbf{p} \\
        \cmidrule(lr){2-3}\cmidrule(lr){4-7}\cmidrule(l){8-8}
        \textbf{Method} & EXC & EXF\textsubscript{1} & NoSub & Empty & Struct. & Value & vs. SAG \\
        \midrule
        \multicolumn{8}{@{}l}{\emph{Cumulative mechanism ladder (full panel, 1,210 tasks)}} \\
        Path Card & 31.3 & 35.0 & 4.7 & 5.5 & 2.3 & 56.2 & $<0.001$ \\
        Gated Repair & 34.8 & 38.5 & 0.0 & 1.2 & 1.8 & 62.1 & 0.353 \\
        Value/Bisection Repair & 33.6 & 36.8 & 0.2 & 0.4 & 1.7 & 64.2 & 0.027 \\
        \textbf{SAG (Ours)} & \textbf{35.8} & \textbf{40.5} & 0.0 & 0.5 & 2.2 & 61.5 & -- \\
        \midrule
        \multicolumn{8}{@{}l}{\emph{Component knockouts\tnote{a}}} \\
        SAG (same batch) & 41.8 & 43.0 & 0.0 & 0.0 & 1.4 & 56.8 & -- \\
        w/o alignment gate & 45.0 & 47.1 & 0.0 & 0.5 & 0.9 & 53.6 & -- \\
        w/o value witnesses & 44.1 & 46.4 & 0.0 & 0.9 & 1.9 & 53.2 & -- \\
        w/o bisection feedback & 45.5 & 48.0 & 0.0 & 0.9 & 1.4 & 52.3 & -- \\
        top-level card & 6.4 & 7.2 & 0.9 & 29.5 & 0.5 & 62.7 & -- \\
        w/o dynamic-key collapse & 37.7 & 39.0 & 0.0 & 0.0 & 0.9 & 61.4 & -- \\
        \bottomrule
      \end{tabular}
    }
    \begin{tablenotes}\footnotesize
    \item[a] Each knockout removes exactly one mechanism from the full solver, measured on \texttt{financial}$+$\texttt{superhero}; the reference row is the same-batch full solver.
    \end{tablenotes}
    \end{threeparttable}
\vspace{-10pt}
\end{table}

The ladder reads as the construction of the solver. \textbf{The induced card alone already beats every baseline.} A single ungated decode conditioned on the data-induced lattice reaches 31.3\% EXC, above the 50-step interactive agent (30.8\%) and every one-shot channel, at one model call per task. In a schema-less world, structure discovery behaves like perception: induced completely from stored documents up front, it is worth more than an entire exploration loop re-deriving it per task. \textbf{Gating and repair convert the remaining mechanical tax into accuracy}, adding +3.5 points (the largest single rung) while eliminating no-submissions and cutting empty executions from 5.5\% to 1.2\%. The gate does this without endangering recall: it accepts all 1,210 gold pipelines across the 11 databases with zero false positives, so by construction it can only remove inadmissible candidates, never a gold-equivalent answer. \textbf{The value contract needs consistency to pay off.} Enforced under a single final attempt (Value/Bisection Repair), it is the one non-monotone rung (33.6\%), because a stricter contract applied to a single derivation occasionally repairs a near-miss into a different wrong program with no independent derivation to fall back on. Full SAG resolves exactly this tension: $k{=}3$ result-space clustering gives the contract multiple independent derivations to arbitrate among, recovering +2.2 points ($p{=}0.027$) and finishing +4.5 above the card alone ($p{<}0.001$). The closest rung, Gated Repair, is not separable from full SAG at pooled significance ($p{=}0.353$); its residual gap concentrates on individual DataWorlds (+5.4 points on \texttt{toxicology}, +4.6 on \texttt{formula\_1}) and is averaged down by pooled testing.

\textbf{Component knockouts confirm where the mechanism lives.} The lower block of Table~\ref{tab:ablation_study} removes exactly one mechanism at a time from the full solver. The card-construction knockouts are decisive: a top-level-fields card collapses the solver from 41.8\% to 6.4\% EXC with 29.5\% empty executions, reproducing the Schema-Direct failure profile because nested and dynamic-key paths vanish from the hypothesis space, and disabling dynamic-key collapse, which floods the card with concrete stored keys until the entry cap truncates coverage, costs 4.1 points. Both card knockouts also behave consistently across the two databases (6.4\% on each for the top-level card; 47.3\% and 28.2\% without collapse), so schema-as-data grounding stands or falls with the completeness of the induced lattice, and collapse is what keeps completeness affordable inside a bounded card. The three guard knockouts (gate, value witnesses, bisection feedback) each shift the subset mean by at most 3.7 points (at most eight records), within run-to-run variation at this scale; the same full configuration scores 33.6\% on \texttt{superhero} in the canonical panel but 29.1\% in the knockout batch under $k{=}3$ sampling, and the guards overlap by construction, since an unwitnessed literal that the value contract would flag typically also drives an empty execution that bisection then localizes. Their measured contribution is panel-level rather than subset-mean: zero mechanical failures in Table~\ref{tab:results}, the zero-false-positive safety of the gate, and the contract that the consistency stage needs, as the following control shows.

\textbf{Consistency is a selection rule, not a sampling bonus.} Wrapping the strongest one-shot baseline in exactly SAG's $k{=}3$ result-clustering rule (on the \texttt{financial}/\texttt{superhero}/\texttt{toxicology} subset) is flat to negative, 36.0\% to 35.5\% mean EXC, and on \texttt{superhero} it amplifies a systematic error (20.9\% to 14.5\%): without admissibility gating, result-space voting arbitrates among unconstrained guesses and selects the consistent-but-wrong cluster. On the same three databases full SAG reaches 46.3\% EXC, +10.8 points over the compute-matched wrap. Consistency pays only on top of gated, witness-grounded candidates; it is a selection rule over admissible programs, not a substitute for grounding.

\subsection{Case Study}
\label{sec:case_study}

\begin{table*}[!t]
    \centering
    \caption{Case study on the \texttt{thrombosis\_prediction} DataWorld.}
    \label{tab:case_study}
    \footnotesize
    \setlength{\tabcolsep}{5pt}
    \renewcommand{\arraystretch}{1.1}
    \newcommand{\channel}[1]{{\scriptsize\itshape\color{commentcolor}#1}}
    \newcommand{\why}[2]{ {\normalfont\itshape\scriptsize\color{#1}$\leftarrow$ #2}}
    \newcommand{\stg}[1]{{\scriptsize\ttfamily\hangindent=11pt\hangafter=1 #1\par}}
    \newcommand{\op}[1]{\textcolor{keywordcolor}{#1}}
    \begin{tabular}{@{}>{\raggedright\arraybackslash}p{0.085\textwidth}>{\raggedright\arraybackslash}p{0.60\textwidth}>{\raggedright\arraybackslash}p{0.215\textwidth}@{}}
        \toprule
        \multicolumn{3}{@{}p{0.97\textwidth}@{}}{%
            \textbf{NLQ.}~\emph{Show the top 100 measurement-code groups that have readings \textbf{for 1987}. Group by supplemental reference range and year, count documents in each group, and sort primarily by document count descending.}}\\[3pt]
        \multicolumn{3}{@{}p{0.97\textwidth}@{}}{%
            \textbf{Witness structure.}~\texttt{measurement\_code\_bags} (25 docs): \texttt{\{ code, bag\_state: \{ supplemental\_reference\_range, $\cdots$ \}, values\_by\_year: \{ "1981": \{ reading\_count, readings: [$\cdots$] \}, "1982": $\cdots$ \} \}}. Per-year readings sit under \emph{dynamic year keys} ($\approx$12 keys per document); only some documents carry a \texttt{"1987"} entry, and the reference range is stored as an explicit \texttt{"missing"} sentinel.}\\
        \midrule
        \textbf{System} & \textbf{Generated pipeline (verbatim, one stage per line)} & \textbf{Executed result / EXC} \\
        \midrule
        \rowcolor{black!4}
        \makecell[tl]{\textbf{Gold}\\ \channel{reference}} &
        \stg{\{ \op{\$addFields}: \{ category: \{ \$ifNull: ["\$bag\_state.\allowbreak supplemental\_reference\_range", "missing"] \} \} \}}%
        \stg{\{ \op{\$project}: \{ dynamic\_entries: \{ \$objectToArray: "\$values\_by\_year" \} \} \}}%
        \stg{\{ \op{\$unwind}: "\$dynamic\_entries" \}}%
        \stg{\textcolor{resultcolor}{\{ \$match: \{ "dynamic\_entries.k": "1987" \} \}}\why{resultcolor}{the 1987 restriction is applied before grouping}}%
        \stg{\{ \op{\$group}: \{ \_id: \{ category, entry\_key \}, document\_count: \{ \$sum: 1 \} \} \}}%
        \stg{\{ \op{\$project}: $\cdots$ \} \{ \op{\$sort}: \{ document\_count: -1, $\cdots$ \} \} \{ \op{\$limit}: 100 \}} &
        \textbf{1 group:}\par {\scriptsize\ttfamily [\{ document\_count: 12, category: "missing", entry\_key: "1987" \}]}\par {\scriptsize\itshape the reference answer} \\
        \midrule
        \makecell[tl]{\textbf{SQL Pivot}\\ \channel{sampled docs}\\ \channel{$\to$ SQL sketch}\\ \channel{$\to$ MQL}} &
        \stg{\{ \op{\$match}: \{ "values\_by\_year.1987": \{ \$exists: true \} \} \}\why{commentcolor}{existence check only, not a restriction}}%
        \stg{\{ \op{\$addFields}: \{ year\_entries: \{ \$objectToArray: "\$values\_by\_year" \} \} \}}%
        \stg{\textcolor{errorcolor}{\{ \$unwind: "\$year\_entries" \}}\why{errorcolor}{all 19 year keys re-expanded; the 1987 restriction is lost}}%
        \stg{\{ \op{\$group}: \{ \_id: \{ supplemental\_reference\_range, \textcolor{errorcolor}{year: "\$year\_entries.k"} \}, doc\_count: \{ \$sum: 1 \} \} \}\why{errorcolor}{every stored year becomes a group; the compound \_id leaks}}%
        \stg{\{ \op{\$sort}: \{ doc\_count: -1 \} \} \{ \op{\$limit}: 100 \}} &
        \xmark~\textbf{19 groups:}\par {\scriptsize\ttfamily [\{ \_id: \{ supplemental\_\allowbreak reference\_range: "missing", year: "1988" \}, doc\_count: 12 \}, \{ \_id: \{ $\cdots$, year: "1987" \}, doc\_count: 12 \}, $\cdots$]}\par {\scriptsize\itshape one group per stored year; plausible rows, wrong content} \\
        \midrule
        \makecell[tl]{\textbf{ReAct}\\ \channel{collection list}\\ \channel{+ interactive}\\ \channel{\phantom{+ }probing}} &
        \stg{\{ \op{\$match}: \{ "values\_by\_year.1987": \{ \$exists: true \} \} \}}%
        \stg{\{ \op{\$project}: \{ range: "\$bag\_state.\allowbreak supplemental\_reference\_range", years: \{ \$objectToArray: "\$values\_by\_year" \} \} \}}%
        \stg{\textcolor{errorcolor}{\{ \$unwind: "\$years" \}}\why{errorcolor}{probing found the map, then the same relational habit}}%
        \stg{\{ \op{\$group}: \{ \_id: \{ range, \textcolor{errorcolor}{year: "\$years.k"} \}, count: \{ \$sum: 1 \} \} \}}%
        \stg{\{ \op{\$sort}: \{ count: -1, $\cdots$ \} \} \{ \op{\$limit}: 100 \} \{ \op{\$project}: \{ \_id: 0, $\cdots$ \} \}} &
        \xmark~\textbf{19 groups:}\par {\scriptsize\ttfamily [\{ supplemental\_\allowbreak reference\_range: "missing", year: "1987", count: 12 \}, \{ $\cdots$, year: "1988", count: 12 \}, $\cdots$]}\par {\scriptsize\itshape same failure as SQL Pivot} \\
        \midrule
        \rowcolor{resultcolor!7}
        \makecell[tl]{\textbf{SAG (ours)}\\ \channel{path card}\\ \channel{+ value}\\ \channel{\phantom{+ }witnesses}} &
        \stg{\textcolor{resultcolor}{\{ \$match: \{ "values\_by\_year.1987.\allowbreak reading\_count": \{ \$gt: 0 \} \} \}}\why{resultcolor}{the literal 1987 is addressed as a stored map key; no flattening at all}}%
        \stg{\{ \op{\$group}: \{ \_id: \{ reference\_range, year: "1987" \}, count: \{ \$sum: 1 \} \} \}}%
        \stg{\{ \op{\$sort}: \{ count: -1 \} \} \{ \op{\$limit}: 100 \} \{ \op{\$project}: \{ \_id: 0, $\cdots$ \} \}} &
        \cmark~\textbf{1 group:}\par {\scriptsize\ttfamily [\{ count: 12, reference\_range: "missing", year: "1987" \}]}\par {\scriptsize\itshape matches the gold result exactly} \\
        \bottomrule
    \end{tabular}
    \par\vspace{3pt}
    \begin{minipage}{0.97\textwidth}
    \scriptsize\raggedright
    \textit{Notes.} Verbatim pipelines generated for the same NLQ, one
    stage per line (only tail boilerplate is elided, $\cdots$).
    \textcolor{errorcolor}{Red} marks the decisive wrong stage and
    \textcolor{resultcolor}{green} the decisive correct one, with the
    reason annotated inline ($\leftarrow$). All four pipelines execute
    without error; the outcome column reports the executed result and
    the EXC verdict (\cmark/\xmark).
    \end{minipage}
\end{table*}

Table~\ref{tab:case_study} contrasts the pipelines generated for one task over the \texttt{thrombosis\_prediction} DataWorld, where per-year readings sit in a dynamic-key map \texttt{values\_by\_year} and the NLQ asks for groups with readings \emph{for 1987}. Both baselines discover the map; \textbf{SQL Pivot} checks \texttt{values\_by\_year.1987} for existence and \textbf{ReAct} probes the collection interactively. Both then fall back on the same relational habit: flatten the map with \texttt{\$objectToArray} and \texttt{\$unwind}, group by the re-expanded year column, and silently drop the 1987 restriction, returning all 19 (range, year) groups in place of the single gold group. The queries parse, execute, and look plausible; only execution-result comparison exposes them. \textbf{SAG} instead resolves the literal ``1987'' through its value index to a witnessed dynamic-map key, addresses \texttt{values\_by\_year.1987} directly, and matches the gold result exactly. The case illustrates the failure mode that separates systems on TEND: not failing to \emph{find} MongoDB-native structure, but reasoning about it relationally once found. SAG's card renders the map as a first-class hypothesis (\texttt{values\_by\_year.<*>}), its witnesses bind literals to stored keys, and its gate rejects unwitnessed paths and literal forms, keeping the pipeline inside the document-native representation of the constraint.

\subsection{Analysis of Relational-Pivot Transfer}
SQL Pivot isolates the relational intermediate under a controlled channel: it shares its sampled-document input with Sampled-doc Direct and self-generates a SQL sketch before translating it to MQL, with no gold SQL, relational catalog, external converter \cite{site24x7,javainuse,RusselSQLtoMongoDB}, or execution feedback, so the pair measures representation mismatch rather than converter coverage. \textbf{The intermediate is a net loss}: 25.6\% versus 28.6\% EXC, with more empty executions (8.8\% vs.\ 7.2\%). The losses concentrate exactly where document structure is query-bearing (Table~\ref{tab:slices}): 14.1\% on the medium anti-transfer slice and 0.0\% on attribute bags, the weakest value-bearing cells in the panel. The mechanism is the one the case study shows verbatim: the sketch expresses document evidence as imagined tables, joins, and globally named columns, and the flatten-then-group translation silently drops dynamic-key restrictions and leaks compound group keys. A self-generated relational intermediate is not merely unhelpful on TEND; it is actively harmful, which is precisely the anti-SQL-transfer property the benchmark is designed to measure.

\section{Related Work}
\label{sec:related_work}

Our work intersects NoSQL data management, natural-language query generation, and execution-centered benchmark design. We position Text-to-NoSQL as a broader problem family whose MongoDB-native instantiation requires executable query synthesis over schema-less document stores.

\subsection{NoSQL Databases}
Document stores such as MongoDB replace fixed table-column catalogs with collections of nested documents, embedded arrays, optional fields, and evolving key sets, which weakens the static schema assumptions of most natural-language database benchmarks. Database research on NoSQL systems has studied scalability and reliability in distributed environments \cite{10.1145/1978915.1978919}, query processing for large-scale data management \cite{MAHAJAN2019120}, and security and privacy \cite{10.1109/TrustCom.2011.70}; these works address system-level concerns. TEND is complementary: it studies how natural-language systems should generate executable MQL when the relevant schema is partly implicit in the document instances, so a benchmark must test inference of field paths, array structure, dynamic-key usage, and result contracts from a concrete database world, not only MongoDB syntax.

A small set of MongoDB-querying resources is better viewed as SQL-to-MQL transfer context, where MongoDB programs are derived from relational schemas and gold SQL. DocSpider \cite{ozer2025docspider} is the most direct example: it migrates the Spider databases into MongoDB with one collection per table and flat documents, then uses LLMs to translate Spider's gold SQL into 4,663 MQL queries gated by execution-equivalence checks. The resulting databases contain no nested, optional, polymorphic, or dynamic-key structure (its own nested-query evaluation falls back to a separate 20-query sample over an external Airbnb collection), so it measures SQL-to-MQL syntax transfer over relational mirrors (Table~\ref{tab:benchmark-comparison}). TEND's claim is complementary and concerns MongoDB-native DataWorld design, where collection boundaries, embedded records, nesting, dynamic keys, optional paths, sparse fields, polymorphic shapes, and witness values are benchmark variables.

Recent NL-to-NoSQL work has also begun to examine language coverage: MultiTEND \cite{qin-etal-2025-multitend} studies multilingual natural-language-to-NoSQL translation with multilingual linking mechanisms, a direction complementary to TEND's MongoDB-native DataWorld focus.

\subsection{Text-to-SQL}
Text-to-SQL connects natural language with relational database systems and has been studied extensively in both the database and NLP communities, from rule-based and sequence-to-sequence methods \cite{10.14778/3407790.3407858,DBS-078,10.1145/3318464.3389776,qi-etal-2022-rasat,popescu-etal-2022-addressing,10.5555/3304222.3304323}, through attention, graph representations, and constrained decoding \cite{10191914,hui-etal-2022-s2sql,10.1609/aaai.v37i11.26536,wang-etal-2020-rat,xu-etal-2018-sql,zheng-etal-2022-hie,guo-etal-2019-towards,10.1609/aaai.v37i11.26535,scholak-etal-2021-picard,10.1145/3534678.3539305}, to the current LLM-centric paradigm \cite{10.14778/3641204.3641221,NEURIPS2023_72223cc6,dong2023c3zeroshottexttosqlchatgpt,10.1145/3654930,nan-etal-2023-enhancing,ren2024purplemakinglargelanguage,10.1145/3589292}, with benchmarks such as WikiSQL \cite{zhong2017seq2sql}, Spider \cite{yu-etal-2018-spider}, KaggleDBQA \cite{lee-etal-2021-kaggledbqa}, and BIRD \cite{NEURIPS2023_83fc8fab} making cross-domain relational evaluation standard. Text-to-SQL remains an important baseline family for Text-to-NoSQL because SQL-derived transfer exposes how much relational structure carries over. However, relational benchmarks rely on static, globally defined schemas, whereas document-store Text-to-NoSQL requires reasoning over nested arrays, sparse fields, dynamic keys, optional paths, and polymorphic structures; relational-pivot and SQL-to-NoSQL channels are therefore useful comparisons but cannot define MongoDB-native benchmark semantics.

\subsection{Natural Language Interfaces for Data Systems}
Natural-language interfaces to data systems extend beyond relational SQL: query generation for graph databases and the Overpass geographic database \cite{10.1007/s10796-022-10295-0,staniek-etal-2024-text}, speech-driven querying \cite{song2024speech,song2022voicequerysystem}, and broader NLIDS surveys \cite{zcan2020StateOT} motivate non-relational access, but these data models do not test schema-less document reasoning over executable aggregation pipelines. Adjacent visualization-facing interfaces study robust, free-form, conversational, and multi-agent text-to-visualization \cite{11112835,10597776,11112929,10.1007/s00778-025-00954-4,10.1145/3786670,10.1145/3788853.3801584}; they share the goal of lowering data-access barriers, but their target artifacts are visualizations or answers over them rather than executable document-store queries. SM3-Text-to-Query \cite{sivasubramaniam2024sm3} is the strongest multi-model resource, representing synthetic medical data across PostgreSQL, MongoDB, Neo4j, and GraphDB and evaluating SQL, MQL, Cypher, and SPARQL; we treat it as complementary, since its contribution is cross-model comparison over a controlled source, whereas TEND isolates MongoDB-native benchmark design with expert-authored DataWorlds, schema-less query-bearing structure, frozen witness execution, and program-diversity control.

\section{Conclusion}
\label{sec:conclusion}
This study formalizes the document-oriented branch of Text-to-NoSQL as executable query synthesis over schema-less MongoDB document stores. We develop TEND, an execution-verified benchmark that makes document structure, dynamic keys, sparse paths, and witness values part of the benchmark semantics; SAG, a schema-as-data grounding solver that induces structure from witness documents before bounded generation and repair; and an execution-centered evaluation protocol with EXC, EXF\textsubscript{1}, and a mutually exclusive outcome decomposition for failure attribution. Extending the grounding index with targeted probes for polymorphic variants and event streams, where agentic exploration remains competitive, is the most direct avenue for further gains. TEND, the SAG implementation, and the evaluation tooling are publicly released to support reproducible progress on Text-to-NoSQL. We hope these contributions together establish schema-less document reasoning as a measurable and improvable dimension of natural-language data access.

\section*{Acknowledgment}
We sincerely thank the anonymous reviewers for their valuable comments and constructive suggestions, which helped improve this paper.

\section*{AI-Generated Content Acknowledgement}
AI coding assistants (Anthropic's Claude, accessed through the Claude Code agent) were used to help implement the experiment code of this work, including the evaluation harness, the baseline and ablation runners, and the result-aggregation scripts described in Section~\ref{sec:exp_analysis}. All AI-assisted code was reviewed, tested, and validated by the authors, and all reported experimental results were produced by executing this validated code against the frozen release data.

\bibliographystyle{IEEEtran}
\bibliography{references}

\end{document}